\documentclass[10pt]{article}
\usepackage{graphicx}

\title{Reactor Neutrino Experiments}
\author{Jun Cao \vspace{0.1in} \\
 {\it Institute of High Energy Physics, P.O. Box 918, Beijing 100049, China} \\
 {\it Email: caoj@ihep.ac.cn}
        }
\date{}

\begin{document}
\maketitle

\begin{abstract}
Precisely measuring $\theta_{13}$ is one of the highest priority in neutrino
oscillation study. Reactor experiments can cleanly determine $\theta_{13}$.
Past reactor neutrino experiments are reviewed and status of next precision
$\theta_{13}$ experiments are presented. Daya Bay is designed to measure
$\sin^22\theta_{13}$ to better than 0.01 and Double Chooz and RENO are designed
to measure it to 0.02-0.03. All are heading to full operation in 2010. Recent
improvements in neutrino moment measurement are also briefed.
\end{abstract}

\section{Introduction}
Reactor anti-neutrino experiments have played a critical role in the history of
neutrinos. Among them, Savannah River Experiment~\cite{reines} by Reines and
Cowan in 1956 observed the first neutrino. Chooz~\cite{chooz} determined the
most stringent upper limit of the last unknown neutrino mixing angle
$\sin^22\theta_{13}<0.17$ in 1998. KamLAND~\cite{kamland} observed the first
reactor anti-neutrino disappearance in 2003 and precisely determined the
$\Delta m_{12}^2=7.58^{+0.21}_{-0.20}\times 10^{-5}eV^2$ recently
\cite{heeger}. Now reactor neutrino experiments become prominent again for
measuring mixing angle $\theta_{13}$ precisely.

Savannah River experiment located 11 m from the reactor core and 12 m
underground. The neutrino target is 200 liters  CdCl$_2$ water solution in two
tanks, sandwiched by 3 liquid scintillator (LS) layers which contained 110
5-inch photomultipliers (PMT). A reactor neutrino interact with a proton in the
target via inverse beta decay (IBD), produce a positron and a neutron,
\begin{equation}
\bar{\nu}_e +p \rightarrow e^+ + n \,.
\end{equation}
The positron produces a prompt signal and the neutron forms a delayed signal
after thermalization and capture on cadmium. The two-fold coincidence can
greatly suppress the background, which is the key to the success. In a modern
experiment to precisely determine $\theta_{13}$, The principle of detecting
reactor neutrinos changes very little since Savannah River Experiment. Large
volume of proton-rich liquid scintillator is used as the neutrino target.
Gadolinium is doped into the liquid scintillator to capture neutron. Gammas of
total energy $\sim 8$ MeV from the neutron capture can cleanly distinguish the
signal from natural radioactivities. Finally, the experiment should locate deep
underground to suppress the cosmogenic backgrounds.

\begin{figure}[!htb]
\begin{center}
\includegraphics[width=0.4\textwidth]{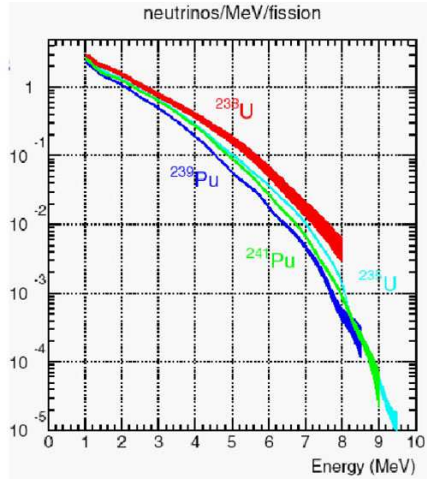}
\caption{Energy spectra of reactor neutrinos.\label{fig:spectra} }
\end{center}
\end{figure}

\begin{figure}[!htb]
\begin{center}
\includegraphics[width=0.4\textwidth]{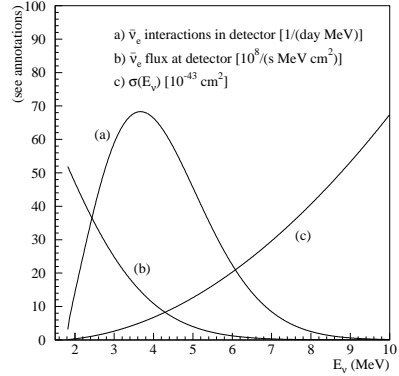}
\caption{Reactor neutrino flux, inverse beta decay cross section, and neutrino
interaction spectrum at a detector based on such reaction.\label{fig:vspectrum}
}
\end{center}
\end{figure}

The electron antineutrino emitted by reactors are dominated by 4 isotopes,
$^{235}$U, $^{239}$Pu, $^{241}$Pu, and $^{238}$U. Other isotopes contribute
only at 0.1\% level. The energy spectra of the first 3 isotopes are measured by
ILL~\cite{ill} while the $^{238}$U spectrum are calculated theoretically, as
shown in Fig.~\ref{fig:spectra}.
The cross section of IBD is calculated to $\sim$ 0.2\% precision with 1/M
corrections~\cite{vogel}. The energy spectrum of reactor neutrinos observed in
a detector based on such reaction is shown in Fig.~\ref{fig:vspectrum}, which
is peaked at around 4 MeV.
Normally we can determine the power fluctuation of a reactor core to $<1$\%
precision. Taking the nuclear fuel burn-up evolution into account, the neutrino
flux can be determined to a precision $\sim 2$\%. The spectra is of similar
precision. The counting rate and spectra were verified by Bugey and
Bugey-3~\cite{bugey}. The absolute counting rate and spectra is critical for
past single-detector experiments, but they will be less important for the next
precision $\theta_{13}$ experiments, since the near-far relative measurement
could cancel the reactor-related error to negligible level.

\begin{figure}[!htb]
\begin{center}
\includegraphics[width=0.4\textwidth]{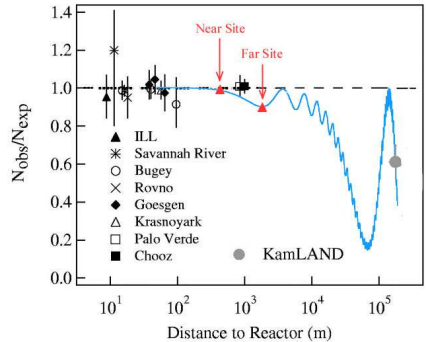}
\caption{Survival probability of reactor antineutrinos of energy 4 MeV,
supposing $\sin^22\theta_{13}=0.1$. Past reactor neutrino experiments are shown
at their baselines.\label{fig:baseline} }
\end{center}
\end{figure}

Neutrino oscillation experiments based on the IBD of reactor antineutrinos are
disappearance experiments. The antineutrino survival probability is
\begin{eqnarray}
P & = & 1- \cos^4\theta_{13} \sin^22\theta_{12} \sin^2\Delta_{21} -
 \cos^2\theta_{12} \sin^22\theta_{13} \sin^2\Delta_{31} -
 \sin^2\theta_{12} \sin^22\theta_{13} \sin^2\Delta_{32} \nonumber\\
 & \approx & 1 - \sin^22\theta_{13}\sin^2\Delta_{31} -
 \sin^22\theta_{12}\sin^2\Delta_{21} \,,
\end{eqnarray}
where $\Delta_{ij}=1.27\Delta m^2_{ij}L/E$, $L$ is the baseline in km, $E$ is
the antineutrino energy in MeV, and $m^2_{ij}$ is the difference of mass square
in eV$^2$. The survival probability is shown in Fig.~\ref{fig:baseline} for
antineutrinos of energy 4 MeV, supposing $\sin^22\theta_{13}=0.1$. The slow
component is due to $\theta_{12}$ oscillation. The first maximum is at around
60 km. The fast component is due to $\theta_{13}$ oscillation, whose first
maximum is at around 2 km. Past reactor neutrino experiments are shown at their
baselines.

\section{Recent Reactor Neutrino Experiments}
There are three modern reactor neutrino experiments, Chooz, Palo Verde, and
KamLAND, that can be used as references of the next precision $\theta_{13}$
experiments.

Chooz experiment~\cite{chooz} located at 1050 m from the reactor cores of the
Chooz power plant in France. Two cores have thermal power of 8.5 GW. Chooz, as
well as Palo Verde, is expired by the atmospheric neutrino anomaly. The
baseline is not at the optimal baseline for $\theta_{13}$ measurement, which is
about 2 km. The experiment took data from March 1997 until July 1998. A total
of 2991 neutrino candidates was collected. The measured vs expected ratio,
averaged over the energy spectrum, is $1.01 \pm 2.8\%{\rm (stat)} \pm 2.7
\%{\rm (syst)}$. The upper limit of $\sin^22\theta_{13}$ is set to be 0.14 at
$\Delta m^2=2.5\times10^{-3}$ eV$^2$.

\begin{figure}[!htb]
\begin{center}
\includegraphics[width=0.5\textwidth]{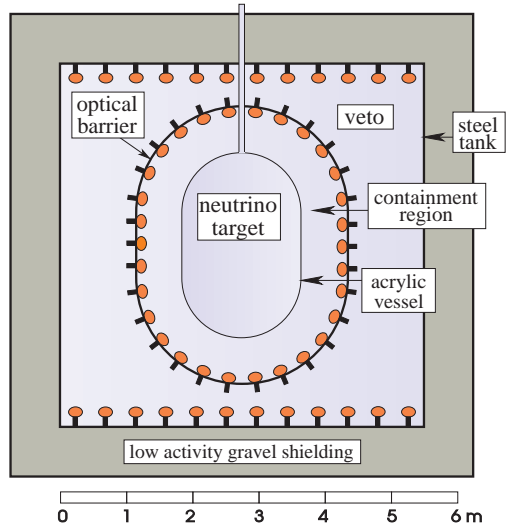}
\caption{Schematic drawing of the Chooz detector.\label{fig:choozdet}}
\end{center}
\end{figure}

The Chooz neutrino detector, schematically shown in Fig.~\ref{fig:choozdet},
was consisted of two layers: a central 5-ton target in a transparent acrylic
container filled with 0.09\% Gd-loaded scintillator (Gd-LS), surrounded by
70-cm thick undoped liquid scintillator, equipped with 192 eight-inch PMTs.
Outside is the veto region, filled with 90-ton undoped liquid scintillator and
equipped with 24 eight-inch PMTs. The detector design, a central volume of
liquid scintillator surrounded by PMTs, common to many other neutrino
experiments such as SNO, Borexino, LSND, as well as later KamLAND, provided a
homogenous detector response. The detector located underground with 300
meter-water-equivalent (m.w.e) overburden. The cosmic muon flux is 0.4
Hz/m$^2$, a factor of 500 reduction from the sea level.

Chooz experiment is an excellent reference for future precision $\theta_{13}$
experiments. The largest drawback of the experiment is the Gd-LS. The
Gd(NO$_3$)$_3$ and hexanol complex was dissolved in aromatic compounds and
paraffin. The complex was not stable enough. The attenuation length dropped
0.42\%/day, or 60\%/year, which significantly degraded the detector performance
and introduced additional detector errors.

Palo Verde experiment~\cite{PaloVerde} was built at the Palo Verde Nuclear
Generating Station, the largest nuclear plant in America. The total thermal
power from three identical pressurized water reactors is 11.6 GWth. Two of the
reactors were located 890 m from the detector, while the third was at 750 m.
Data were collected in the period between October 1998 and July 2000. The
detector located underground with only 32 m.w.e overburden, or a factor of 5
reduction of the cosmic muons. Extra efforts must be made to further reduce
backgrounds. Thus the detector was designed to be segmented to take full
advantage of the triple coincidence given by the $e^+$ ionization and two
annihilation gammas. The detector consisted of 66 tanks filled with 0.1\%
Gd-LS. Each cell was 9 m long, with $12.7 \times 25.4$ cm$^2$ cross section,
viewed by two 5-inch PMTs, one at each end. Total target mass is 12 ton. The
central detector was shielded by 1-m thick water buffer to moderate cosmogenic
neutrons. Outside of the water tanks were 32 large liquid scintillator counters
as the muon veto.

\begin{figure}[!htb]
\begin{center}
\includegraphics[width=0.4\textwidth]{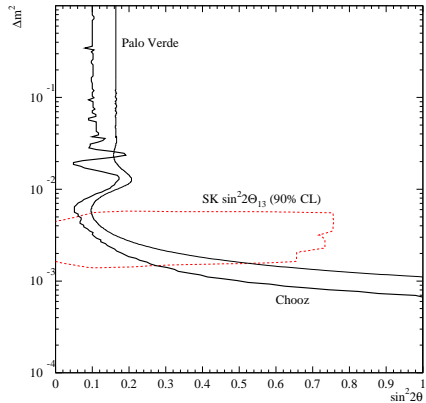}
\caption{Exclusion plot showing the allowed region of $\sin^22\theta_{13}$ and
$\Delta m^2$ based on the Super-Kamiokande preliminary analysis (the region
inside the dotted curve).\label{fig:th13limit}}
\end{center}
\end{figure}

Due to the shallow overburden, Palo Verde suffered from large backgrounds. The
energy averaged measured vs expected ratio is $1.01\pm 2.4\% {\rm (stat)} \pm
5.3\%{\rm (syst)}$. The exclusion curves for $\sin^22\theta_{13}$ of both Chooz
and Palo Verde are shown in Fig.~\ref{fig:th13limit}~\cite{bemporad}.

One highlight of Palo Verde is that stabler Gd-LS is achieved. The Gd-LS use
2-ethyl-hexanoic acid as complex agent of GdCl$_3$. The complex is dissolved
into a cocktail of 36\% pseudocumene, 60\% mineral oil, and 4\% alcohol (use to
increase the solubility of the Gd complex). The attenuation degradation is 12\%
in the first year and 3\% in the second.

KamLAND experiment \cite{kamland} locates in Japan, 500 m from Super-Kamiokande
experiment. The rock overburden is 2700 m.w.e., provide an excellent shielding.
It detects neutrinos from more than 60 reactors in Japan and Korea. The average
baseline is $\sim$ 180 km. The experiment starts from 2002 and is running until
now. The detector has also a two-layer structure. 1000 ton liquid scintillator
(undoped) is suspended with a 135 $\mu$m-thick balloon in 1800 m$^3$ mineral
oil. They are contained in an 18-m diameter stainless steel tank. 1325 17-inch
PMTs and 554 20-inch PMTs installed in mineral oil. Due to the large
backgrounds from the radioactivity accumulated on the balloon, a fiducial
volume cut is applied for analysis, as well as a higher energy threshold. These
two cuts bring 4.7\% and 2.3\% systematic uncertainties and result in a total
systematic uncertainty of 6.5\%.

KamLAND observed the first reactor anti-neutrino disappearance. The energy
averaged measured vs expected ratio is $R=0.658 \pm 0.044{\rm (stat)} \pm 0.047
{\rm (syst)}$. It confirmed antineutrino disappearance at 99.998\% C.L. It
verified the solar neutrino oscillation with man-made neutrino source and
exhibited the oscillation pattern in terms of L/E, thus excluded other
hypothesis such as neutrino decay and decoherence. It also provided the best
$\Delta m^2_{21}$ measurement. Recently KamLAND updated the analysis with 2007
data~\cite{heeger}. With improved analysis, the fiducial volume uncertainty is
reduced to 1.8\% and energy threshold uncertainty to 1.5\%, resulting in a
total systematic uncertainty of 3.4\%. With KamLAND data only, the oscillation
parameter is determined to be
\begin{eqnarray*}
\tan^2\theta_{12} &=& 0.56^{+0.14}_{-0.09} \;,\\
\Delta m^2_{21} &=& 7.58^{+0.21}_{-0.20}\times 10^{-5}\ {\rm eV}^2 \;,
\end{eqnarray*}

\section{Precision $\theta_{13}$ Experiments}
There are six mixing parameters related with neutrino oscillation. We have
measured $|\Delta m^2_{32}|$ and $\sin^22\theta_{23}$ via atmospheric and
accelerator experiments, $|\Delta m^2_{21}|$ and $\sin^22\theta_{12}$ via solar
and reactor experiments. $\sin^22\theta_{13}$, CP violating phase $\delta_{\rm
CP}$, and the sign of $\Delta m^2_{32}$ are still unknown. The upper limit of
$\sin^22\theta_{13}$ is set to be $<0.17$ by Chooz, while the other two mixing
angles are maximal or close to maximal.

Next generation of long baseline accelerator neutrino experiments, with a
proton driver in the megawatt class or above, together with a detector of more
than 100 kilotons mass, will have chance to measure all three unknowns, as long
as $\sin^22\theta_{13}$ is larger than 0.01. However, since all three unknowns
involve in the oscillation, it is hard to clearly determine any of them. A
reactor experiment at medium baseline of around 2 km can determine the
magnitude of $\theta_{13}$ independent of the influence of CP violation and the
mass hierarchy. Meanwhile, $\theta_{13}$ modulate the effect of CP violation.
If $\sin^22\theta_{13}$ is smaller than 0.01, a neutrino factory will be
required. Given these reasons, it is recommended in APS Neutrino
Study~\cite{numatrix} in 2004 that "as a high priority, $\ldots$, An
expeditiously deployed multi-detector reactor experiment with sensitivity to
$\bar{\nu}_e$ disappearance down to $\sin^22\theta_{13}=0.01$."

To improve the $\sin^22\theta_{13}$ sensitivity by an order, both systematic
and statical uncertainties must be greatly reduced from past reactor neutrino
experiments. It is first proposed by Krasnoyarsk~\cite{Krasnoyarsk} to use
near-far detectors to cancel systematic errors with relative measurement. Since
2002, eight experiments are proposed~\cite{whitepaper}. They are Angra in
Brazil, Braidwood in US, Daya Bay in China, Diablo Canyon in US, Double Chooz
in France, KASKA in Japan, Krasnoyarsk in Russia, and RENO in Korea. After
several years' R\&D, four of them are cancelled. Angra is still under R\&D.
Daya Bay, Double Chooz, and RENO are heading to construction and all schedule
the full operation in 2010.

Daya Bay aims at $\sin^22\theta_{13}$ sensitivity better than 0.01 while Double
Chooz and RENO aim at $0.02\sim0.03$ at 90\% C.L. in 3 years. There are many
similarities in the detector design, based on past experiences.
\begin{itemize}
\item Identical detectors are put at the near site(s) and the far site to
cancel out reactor-related uncertainties and part of detector-related
uncertainties.

\item Gadolinium doped liquid scintillator (Gd-LS) is adopted as the neutrino
target and extensive R\&D is carried out to improve the quality of Gd-LS. The
Daya Bay Gd-LS is based on Linear Alkylbenzene (LAB). Solid complex of GdCl$_3$
and TMHA is dissolved in pure LAB, as well as fluors. The Double Chooz uses
$\beta$-diketonates to complex GdCl$_3$. Solid complex is dissolved in a
mixture of 20\% PXE and 80\% dodecane. Both recipes show good long term
stability with small or ton-level samples.

\item Antineutrino detector is designed to be a cylinder with 3 layers,
extended from a two-layer design of Chooz and KamLAND. The inner layer is
Gd-LS, surrounded by a layer of undoped liquid scintillator (LS), called
$\gamma$-catcher, to contain gamma energy. The outer layer is oil buffer to
shield detector and ambient radioactivities. Three kinds of liquid are
separated by optical transparent acrylic vessels. Scintillation light produced
by neutrino reactions are viewed by PMTs installed in the oil buffer. Such a
design will minimize the analysis cuts to reduce the detector related
uncertainties, based on the experience of Chooz and KamLAND.

\item Passive and active detector shielding, besides enough overburden,
is improved comparing to past experiments, which allows a lower threshold to
trigger reactor neutrinos in the full energy range to reduce the systematic
uncertainties.
\end{itemize}

With these improvements, the reactor-related uncertainties could cancel out or
be reduced to 0.1\% level. The detector-related uncertainties could be lowered
to 0.38\% to 0.6\%. And the background-related uncertainties could be
suppressed to 0.4\% to 1\%.

\subsection{Angra}
Angra experiment \cite{angra} locates in the neighborhood of Rio de Janeiro,
Brazil. Two reactors, Angra-I and Angra-II, have thermal power 1.8 GW and 4 GW,
respectively. A initial design for Angra is to put a 50-ton near detector 300 m
away from Angra-II, with 250 m.w.e overburden, and a 500-ton far detector at a
distance of 1500 m, with 2000 m.w.e overburden. A statistical precision of
$\sin^22\theta_{13} <0.006$ at 90\% C.L. could be obtained in three years.
\begin{figure}[!htb]
\begin{center}
\includegraphics[width=0.6\textwidth]{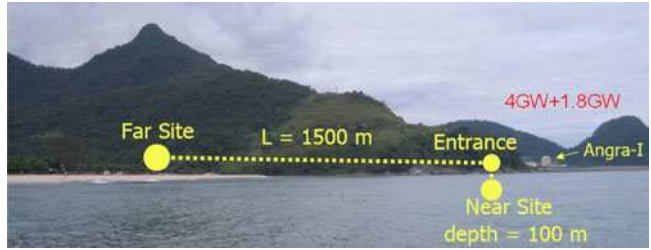}
\caption{Layout of Angra experiment.\label{fig:angra}}
\end{center}
\end{figure}

Angra is planned to start in 2013. The collaboration is consisted of 30
researchers from 11 institutes. The Brazilian group also participates the
Double Chooz experiment. In March 2007, a very near (prototype) detector for
safeguards study was approved.

\subsection{Daya Bay}

Daya Bay experiment \cite{dyb} is the only one that is designed to measure
$\sin^22\theta_{13}$ to better than 0.01 at 90\% C.L. in near future.

\begin{figure}[!htb]
\begin{center}
\includegraphics[width=0.5\textwidth]{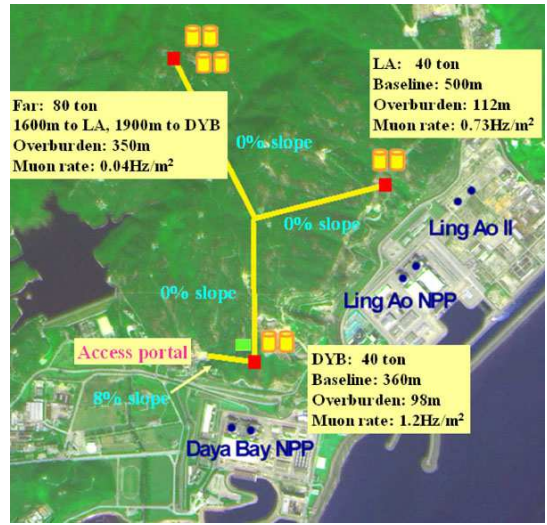}
\caption{Layout of the Daya Bay experiment.\label{fig:dyb}}
\end{center}
\end{figure}

Daya Bay Nuclear Power Plant (NPP) and LingAo NPP locate in the south of China,
55 km to the northeast of Hong Kong and 45 km to the east of Shen Zhen city.
The two NPPs are about 1100 m apart. Each NPP has two cores running. Another
two cores, called LingAo II, are expected to commission in 2010 and 2011,
respectively. The thermal power of each core is 2.9 GW. Hence the existing
total thermal power is 11.6 GW, and will be 17.4 GW starting from 2011. The
NPPs are adjacent to high hills that can provide protection from cosmic rays.
The rock is hard granite of density around 2.7 g/cm$^3$, ideal for tunnel
construction. A horizontal tunnel will be built to connect the underground
experimental halls, which makes the logistics much easier and large detectors
could be built. The experiment layout is shown in Fig.~\ref{fig:dyb}. The
experiment is consisted of three experimental sites, the Daya Bay near site
(DYB), the LingAo near site (LA), and the far site. The baseline, overburden,
and predicted muon rate for each site are also shown in the figure.

To measure $\sin^22\theta_{13}$ down to 0.01, systematic uncertainties should
be $<$0.5\%. Cross check with redundancy and extra handling on systematic
uncertainties are necessary for such a high precision measurement. The
experiment is designed to have multiple antineutrino detectors (ADs) at each
site, two at each near site and four at the far site. The detector systematic
errors can be cross checked to 0.1\% level in 3 years via side-by-side
calibration of two detectors at the same near sites. After the 1st 3 years'
data taking, ADs will be swapped between near and far sites to change
systematics and further improve the experimental sensitivity. All ADs will be
assembled in a cleaning room in Surface Assembly Building (SAB) and be filled
in an underground Liquid Scintillator Hall with the same batch of Gd-LS and LS,
and with a reference tank for precise target mass measuring.

\begin{figure}[!htb]
\begin{center}
\includegraphics[width=0.5\textwidth]{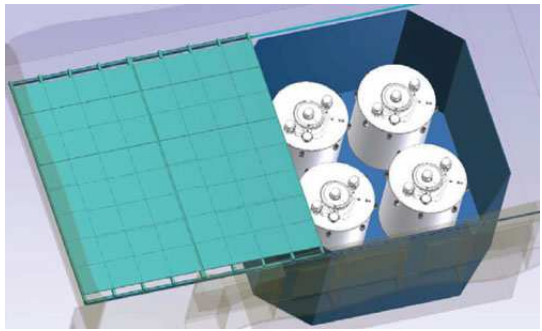}
\caption{Four antineutrino detectors in the water pool of the far hall of Daya
Bay. Above the pool is the RPC detector on rails. \label{fig:dyb_muon}}
\end{center}
\end{figure}

The ADs will be submerged in a big water pool, with at least 2.5 m water
shielding in any direction, to shield the cosmogenic neutrons produced in the
rock, as well as radioactivities from the rock and the air. The pool is filled
with high purity water. Around 300 8-in PMTs are instrumented in each pool to
form a muon cherenkov detector. The pool is divided into an inner layer and an
outer layer optically, thus muon efficiency can be cross checked and neutron
rejection power can be studied. The water pool is covered with another muon
detector, Resistive Plate Chambers (RPCs). 4 layers RPCs with alternating x and
y strips are assembled in 2m$\times$2m modules and supported by a steel frame
on rails. The RPCs can be slided to aside for AD installation and maintenance.
Combining the water cherenkov detectors and RPCs, muon efficiency will be
determined to 99.5$\pm0.25$\%. Together with the overburden of the Daya Bay
sites, the uncertainties of two major backgrounds, fast neutron background and
accidental coincidence background, will be kept below 0.1\%. The scheme of the
muon system is shown in Fig.~\ref{fig:dyb_muon}.

\begin{figure}[!htb]
\begin{center}
\includegraphics[width=0.4\textwidth]{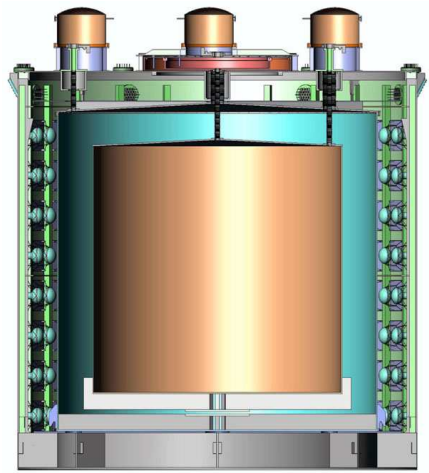}
\caption{The Daya Bay antineutrino detector.\label{fig:dyb_ad}}
\end{center}
\end{figure}

The structure of the AD is shown in Fig.~\ref{fig:dyb_ad}. The inner layer is
20 ton Gd-LS, surrounded by 42.5 cm LS as the $\gamma$-catcher. 192 8-in PMTs
are instrumented in a 50-cm thick oil buffer. To increase the photocathode
coverage and simplify the mechanical structure, two reflective panels are put
at the bottom and the top of the detector, adjacent to the $\gamma$-catcher.
The outer detector tank is made of stainless steel, with dimension
$\phi$5m$\times$5m. Three automatic calibration devices are installed on the
top. The detector is designed to be movable in the tunnel to enable the filling
at the same site and the detector swapping. The total weight of an AD is around
100 ton. The energy resolution is $\sim12\%/\sqrt{E}$ by simulation.

\begin{figure}[!htb]
\begin{center}
\includegraphics[width=0.5\textwidth]{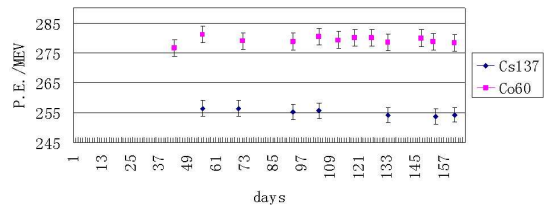}
\caption{Stability monitoring of Daya Bay Gd-LS with IHEP prototype via total
photoelectrons collected for two sources, since Jan. 2007.\label{fig:dyb_ls}}
\end{center}
\end{figure}

The AD design is studied with a $\phi$2m$\times$2m prototype at IHEP, starting
running since Mar. 2006. The prototype has a two-layer structure. A
$\phi$1m$\times$1m acrylic vessel filled with LS is viewed by 45 8-in PMTs in
the outer mineral oil layer. Two reflective panels locate at the bottom and the
top of the prototype. In Jan. 2007, the LS was replaced with 800 liter 0.1\%
Gd-LS. The stability of Gd-LS is monitored in the prototype and no apparent
degradation found, as shown in Fig.~\ref{fig:dyb_ls}.

\begin{figure}[!htb]
\begin{center}
\includegraphics[width=0.4\textwidth]{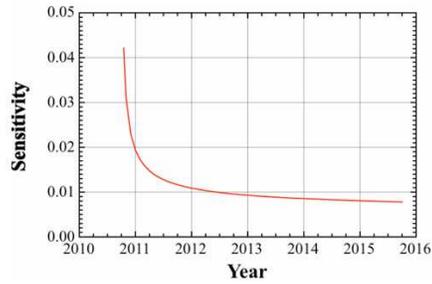}
\caption{Sensitivity of Daya Bay at 90\% C.L. for $\Delta
m^2_{32}=2.5\times10^{-3}$ eV$^2$.\label{fig:dyb_sens}}
\end{center}
\end{figure}

The Daya Bay collaboration is consisted of 34 institutes from China, Czech,
Russia, and United States, about 180 collaborators. The bidding of civil
construction was done recently and the tunnel construction will start in Oct.
2007. The civil construction will take 22 months. The SAB will be ready in Jun.
2008, when the detector assembly can be started. The first near site will start
data taking in Jun. 2009 and full operation will be in Oct. 2010. The expected
sensitivity of Daya Bay versus year is shown in Fig.~\ref{fig:dyb_sens}.

\subsection{Double Chooz}

\begin{figure}[!htb]
\begin{center}
\includegraphics[width=0.5\textwidth]{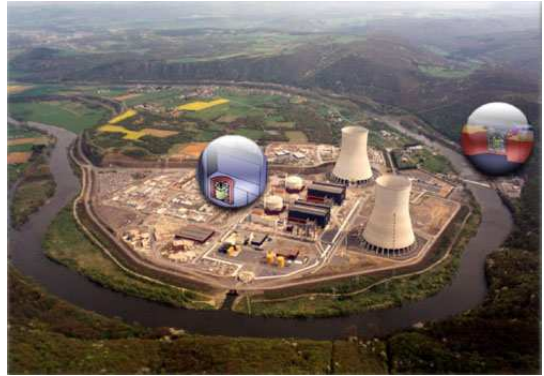}
\caption{Layout of Double Chooz.\label{fig:dchooz}}
\end{center}
\end{figure}

Double Chooz \cite{dchooz} locates in Ardennes, France. It uses the existing
detector site of Chooz experiment as the far site. The baseline is 1050 m and
overburden is 300 m.w.e. The size of the pit available for the detector is
7m$\times$7m. A near site will be constructed $\sim$ 280 m away from the
reactor cores, with $\sim$ 80 m.w.e. overburden. The layout is shown in
Fig.~\ref{fig:dchooz}

\begin{figure}[!htb]
\begin{center}
\includegraphics[width=0.5\textwidth]{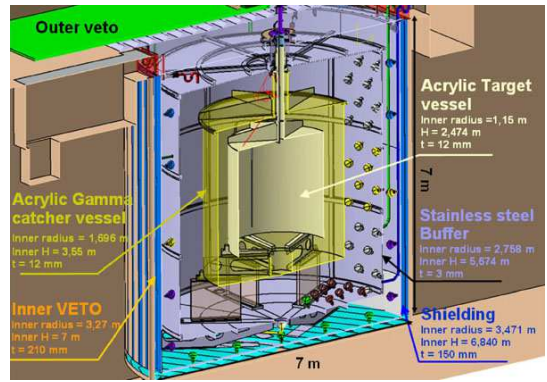}
\caption{Detector of Double Chooz.\label{fig:dchooz_det}}
\end{center}
\end{figure}

Each site has a single neutrino detector of target mass 8.3 ton. The detector
design is shown in Fig.~\ref{fig:dchooz_det}, with dimensions of each volume.
The $\gamma$-catcher thickness is 55 cm and the oil buffer thickness is 105 cm.
The dimension of the neutrino detector is $\phi$2.76m$\times$5.67m. Outside is
a 50 cm veto layer filled with oil and a 15 cm shielding layer made of steel.

\begin{figure}[!htb]
\begin{center}
\includegraphics[width=0.5\textwidth]{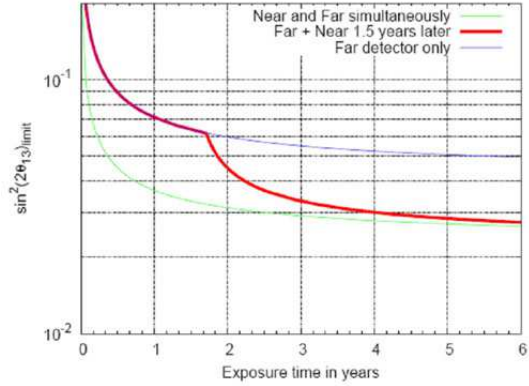}
\caption{Sensitivity of Double Chooz experiment\label{fig:dchooz_sens}}
\end{center}
\end{figure}

Double Chooz is consisted of $\sim$ 100 scientists from 32 institutions. The
experiment has been approved by most of the respective Scientific Councils.
Detector construction and integration will start in 2007. Because the far site
exists, it will take data with a single far site starting from 2008. With the
improved detector design, it will measure $\sin^22\theta_{13}$ to 0.06 at 90\%
C.L. in two years. The near site need civil construction and is expected to
start in 2010. The goal $\sin^22\theta_{13}$ sensitivity is 0.02-0.03, shown in
Fig.~\ref{fig:dchooz_sens}.

\subsection{RENO}
\begin{figure}[!htb]
\begin{center}
\includegraphics[width=0.5\textwidth]{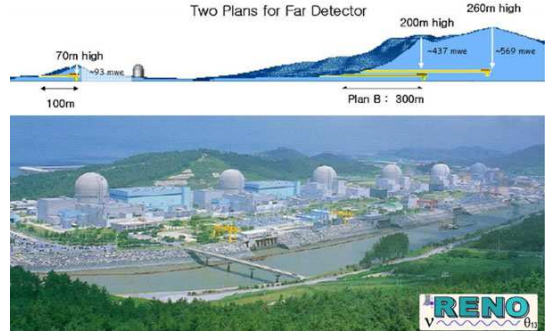}
\caption{Site and layout of RENO.\label{fig:reno}}
\end{center}
\end{figure}

RENO \cite{reno} is a Korean based experiment. The YongGwang NPP has equally
spaced 6 cores with thermal power 16.4 GW. The distance between 2 cores is 256
m. The experiment layout is shown in Fig.~\ref{fig:reno}. The near site will be
located 150 m from the midpoint of 6 cores. The overburden is $\sim$ 93 m.w.e.
The far site is 1500 m from the midpoint, with $\sim$ 437 m.w.e.

\begin{figure}[!htb]
\begin{center}
\includegraphics[width=0.4\textwidth]{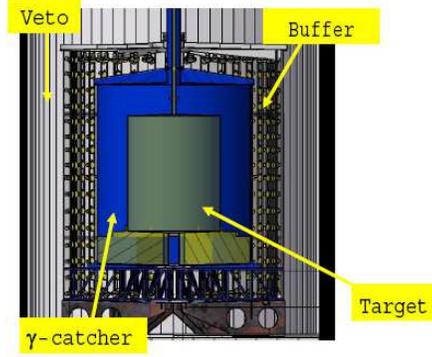}
\caption{The detector design of RENO.\label{fig:reno_det}}
\end{center}
\end{figure}

Each site has a single neutrino detector of target mass 15 ton. The detector
design is shown in Fig.~\ref{fig:reno_det}. The target is a cylinder of
$\phi$2.8m$\times$3.2m. The $\gamma$-catcher thickness is 60 cm and the oil
buffer thickness is 70 cm. 573 8-inch PMTs are implemented in the oil buffer.
The energy resolution is expected to be $7.7\%/\sqrt{E}$. Outside the neutrino
detector is a water cherenkov detector of 1-m thickness.

\begin{figure}[!htb]
\begin{center}
\includegraphics[width=0.4\textwidth]{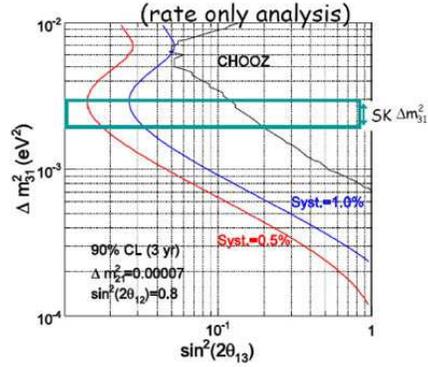}
\caption{Sensitivity of RENO.\label{fig:reno_sens}}
\end{center}
\end{figure}

The RENO collaboration is consisted of 43 collaborators from 11 institutes in
Korea and 2 institutes in Russia. The project was approved in 2005. Geological
survey was completed in May 2007. Detector construction will begin at the end
of 2007. Data taking is expected to start in early 2010. The projected
sensitivity of RENO is shown in Fig.~\ref{fig:reno_sens}, which is $\sim$ 0.02
at 90\% C.L. in 3 years.

\section{Neutrino Magnetic Moments}
\begin{figure}[!htb]
\begin{center}
\includegraphics[width=0.5\textwidth]{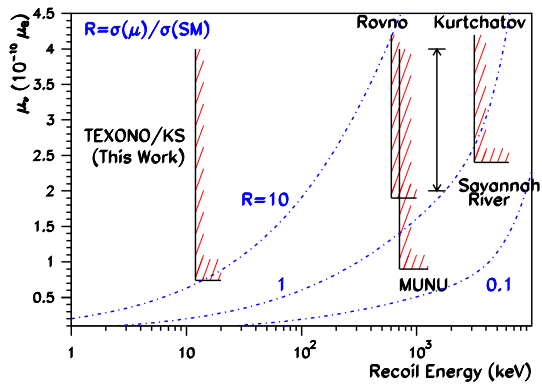}
\caption{Limit of neutrino magnetic moment.\label{fig:texono}}
\end{center}
\end{figure}

The Minimal Standard Model predicts a tiny neutrino moment $\sim
3\times10^{-19}\mu_{\rm B}\cdot m_\nu/1$ eV. Many extensions beyond Standard
Model give larger predictions, say, $10^{-10}-10^{-12}\mu_{\rm B}$ \cite{nmm}.
Reactor neutrinos can be used for direct search of neutrino moments at very
short baseline via neutrino-electron scattering. The contribution to the
scattering cross section is
\begin{equation}
\left(\frac{d\sigma}{dT}\right)_\mu = \frac{\pi\alpha^2_{\rm
em}}{m_e^2}\frac{1-T/E_\nu}{T}\mu_\nu^2 \, ,
\end{equation}
which becomes larger as the recoil energy $T$ goes lower. Thus lowering the
detection threshold to keV level or even lower is desirable. There were limits
from DONUT, LSND, Borexino, Super-K, KamLAND, MUNU, and TEXONO experiments. The
upper limit was $1.0\times10^{-10}\mu_{\rm B}$, established by MUNU, shown in
Fig.~\ref{fig:texono}. Recently TEXONO improved the upper limit to
$\mu_\nu<0.74\times10^{-10}\mu_{\rm B}$ \cite{texono} and GEMMA announced their
first result to be $\mu_\nu<0.58\times10^{-10}\mu_{\rm B}$ \cite{gemma}, both
at 90\% C.L. Both experiments use a single crystal HPGe detector of mass at 1
kg level at a distance 10-20 m from the reactor. The backgrounds are at 1
counts/kg/day/keV level with an energy threshold $\sim$10 keV. TEXONO is taking
data with a 200 kg CsI(Tl) crystal array to first measure SM neutrino-electron
scattering at MeV range. At the same time, it is developing an 1 kg Ultra-Low
Energy Ge detector to lower the threshold to 100 eV. A
$2.0\times10^{-11}\mu_{\rm B}$ limit could be reached. GEMMA aims at the level
of $1.5\times10^{-11}\mu_{\rm B}$ with 4 times larger detector.

\section{Summary}

\indent\par Precisely measuring $\theta_{13}$ is of high priority in neutrino
oscillation study. Sensitivity to $\sin^22\theta_{13} < 0.01$ is achievable
based on experiences of past reactor neutrino experiments. Four $\theta_{13}$
experiments are in progress. Three of them project similar timeline, full
operation starting in 2010, as listed in table~\ref{tab:com}. Double Chooz will
get to 0.06 before 2010 using a single far detector. Upper limit on neutrino
magnetic moment is improved to $0.74\times10^{-10}\mu_{\rm B}$ by TEXONO and
$0.58\times10^{-10}\mu_{\rm B}$ by GEMMA.
\begin{table}[!htb]
\begin{center}
\begin{tabular}{|c|c|c|c|c|}
 \hline
 & Luminosity in 3 year  & Overburden & Projected & Projected full \\
 & (ton$\cdot$GW$\cdot$y) & near/far(mwe) & Sensitivity & operation date \\
 \hline
 Daya Bay & 4200 & 270/950 & $<$0.01 & End of 2010 \\ \hline
 Double Chooz & 210 & 80/300 & 0.02-0.03 & 2010 \\ \hline
 RENO & 740 & 90/440 & $\sim$ 0.02 & Early 2010 \\ \hline
\end{tabular}
 \caption{Summary of three $\theta_{13}$ experiments.\label{tab:com}}
\end{center}
\end{table}


\begin{thebibliography}{99}

\bibitem{reines} C. L. Cowan {\it et al.} Science 124, 103 (1956); F. Reines and
C. L. Cowan, Jr., Nature 178, 446 (1956).

\bibitem{chooz} M. Apollonio {\it et al.}, Eur. Phys. J. C27, 331 (2003)

\bibitem{kamland}  T. Araki {\it et al.}, Phys. Rev. Lett. {\bf 94}, 081801 (2005).

\bibitem{heeger} K. Heeger, talk at TAUP2007, http://www.awa.tohoku.ac.jp/taup2007/

\bibitem{ill} H.~Kwon {\it et al.}, Phys.\ Rev.\ D{\bf 24}, 1097 (1981).

\bibitem{vogel} P. Vogel and J. F. Beacom, Phys. Rev. {\bf D}60, 053003 (1999);
A. Kurylov {\it et al.}, Phys. Rev. {\bf C}67, 035502 (2003).



\bibitem{bugey} Y.~Declais {\it et al.}, Phys.\ Lett.\ {\bf B338}, 383 (1994).
B.~Ackar {\it et al.}, Nucl.\ Phys.\ {\bf B434}, 503 (1995); B. Ackar {\it et
al.}, Phys.\ Lett.\ {\bf B374}, 243 (1996).

\bibitem{PaloVerde}
F.~Boehm {\it et al.}, Phys.\ Rev.\ D{\bf 62}, 072002 (2000).

\bibitem{bemporad} C. Bemporad, G. Gratta, P. Vogel, Rev. Mod. Phys. 74, 297 (2002).

\bibitem{numatrix} The Neutrino Matrix, physics/0411216.

\bibitem{Krasnoyarsk} V. Martemyanov {\it et al.}, hep-ex/02111070.

\bibitem{whitepaper} $\theta_{13}$ White Paper, hep-ex/0402041.

\bibitem{angra} http://www.e-science.unicamp.br/angra/

\bibitem{dyb} http://dayabay.ihep.ac.cn/;
Daya Bay Proposal, hep-ex/0701029

\bibitem{dchooz} http://doublechooz.in2p3.fr/;
Double Chooz Proposal, hep-ex/0606025

\bibitem{reno} http://neutrino.snu.ac.kr/RENO/

\bibitem{nmm} H.T. Wong and H.B. Li, Mod.
Phys. Let. {\bf A}20, 1103 (2005); A. B. Balantekin, hep-ph/0601113.

\bibitem{texono} TEXONO Collaboration, Phys. Rev. {\bf D}75, 012001 (2007).

\bibitem{gemma} A.G. Beda {\it et al.}, ArXiv:0705.4576 [hep-ex].

\end{thebibliography}
\end{document}